\documentclass[aps,prl,twocolumn,superscriptaddress]{revtex4-1}

\usepackage{amsmath}
\usepackage{amssymb}
\usepackage{wasysym}
\usepackage{graphicx}
\usepackage{color}
\usepackage{physics}
\usepackage{siunitx}

\usepackage[english,nomargin,inline,marginclue,draft]{fixme}
\fxusetheme{colorsig}
\FXRegisterAuthor{cg}{acg}{CG}  
\FXRegisterAuthor{ib}{aib}{IB}
\makeatletter
\renewcommand*\FXLayoutInline[3]{%
  {\@fxuseface{inline}\ignorespaces{\color{fx#1}[#3: #2]}}}
\makeatother

\begin{document}

\title{Robust Bilayer Charge-Pumping for \\
Spin- and Density-Resolved Quantum Gas Microscopy}
\author{Joannis Koepsell}
\email{joannis.koepsell@mpq.mpg.de}
\affiliation{Max-Planck-Institut f\"{u}r Quantenoptik, 85748 Garching, Germany}
\affiliation{Munich Center for Quantum Science and Technology (MCQST), 80799 M\"{u}nchen, Germany}
\author{Sarah Hirthe}
\affiliation{Max-Planck-Institut f\"{u}r Quantenoptik, 85748 Garching, Germany}
\affiliation{Munich Center for Quantum Science and Technology (MCQST), 80799 M\"{u}nchen, Germany}
\author{Dominik Bourgund}
\affiliation{Max-Planck-Institut f\"{u}r Quantenoptik, 85748 Garching, Germany}
\affiliation{Munich Center for Quantum Science and Technology (MCQST), 80799 M\"{u}nchen, Germany}
\author{Pimonpan Sompet}
\affiliation{Max-Planck-Institut f\"{u}r Quantenoptik, 85748 Garching, Germany}
\affiliation{Munich Center for Quantum Science and Technology (MCQST), 80799 M\"{u}nchen, Germany}
\author{Jayadev Vijayan}
\affiliation{Max-Planck-Institut f\"{u}r Quantenoptik, 85748 Garching, Germany}
\affiliation{Munich Center for Quantum Science and Technology (MCQST), 80799 M\"{u}nchen, Germany}
\author{Guillaume Salomon}
\affiliation{Max-Planck-Institut f\"{u}r Quantenoptik, 85748 Garching, Germany}
\affiliation{Munich Center for Quantum Science and Technology (MCQST), 80799 M\"{u}nchen, Germany}
\author{Christian Gross}
\affiliation{Max-Planck-Institut f\"{u}r Quantenoptik, 85748 Garching, Germany}
\affiliation{Munich Center for Quantum Science and Technology (MCQST), 80799 M\"{u}nchen, Germany}
\affiliation{Physikalisches Institut, Eberhard Karls Universit\"{a}t T\"{u}bingen, 72076 T\"{u}bingen, Germany}
\author{Immanuel Bloch}
\affiliation{Max-Planck-Institut f\"{u}r Quantenoptik, 85748 Garching, Germany}
\affiliation{Munich Center for Quantum Science and Technology (MCQST), 80799 M\"{u}nchen, Germany}
\affiliation{Fakult\"{a}t f\"{u}r Physik, Ludwig-Maximilians-Universit\"{a}t, 80799 M\"{u}nchen, Germany}

\date{\today}

\begin{abstract}
Quantum gas microscopy has emerged as a powerful new way to probe quantum many-body systems at the microscopic level. However, layered or efficient spin-resolved readout methods have remained scarce as they impose strong demands on the specific atomic species and constrain the simulated lattice geometry and size. Here we present a novel high-fidelity bilayer readout, which can be used for full spin- and density-resolved quantum gas microscopy of two-dimensional systems with arbitrary geometry. Our technique makes use of an initial Stern-Gerlach splitting into adjacent layers of a highly-stable vertical superlattice and subsequent charge pumping to separate the layers by $21\,\mu$m. This separation enables independent high-resolution images of each layer. We benchmark our method by spin- and density-resolving two-dimensional Fermi-Hubbard systems. Our technique furthermore enables the access to advanced entropy engineering schemes, spectroscopic methods or the realization of tunable bilayer systems.
\end{abstract}

\maketitle
Quantum simulation has opened a new and unique window to explore static and dynamical properties of quantum matter \cite{Gross2017, Georgescu2014, Blatt2012, Weimer2010, Devoret2013}, difficult to access with classical numerical computations. Strongly-correlated materials are typically simulated with Fermi-Hubbard systems, in which the intricate interplay between density (charge) and spin degrees of freedom is believed to contain essential ingredients for the physics of high-temperature superconductivity \cite{Lee2006, Keimer2015}. Despite its simple form, the two-dimensional Fermi-Hubbard model still poses major challenges to the exploration of its phase diagram. In addition, it is an open question to what extent the layered structure in real materials like the cuprates affects the resulting physical properties \cite{Damascelli2003, Yuli2008}.

Quantum gas microscopy of two-dimensional Fermi-Hubbard systems promises to shed new light on the  interplay between antiferromagnetic order and mobile density (charge) dopants. Several aspects were recently explored, such as long-range antiferromagnetic correlations \cite{Mazurenko2017}, spin and density transport \cite{Nichols2018, Brown2019a} or the effect of doping on two-point spin-correlations \cite{Chiu2019}. However, most experiments are strongly constrained in their accessible observables. It is usually possible to measure either density observables, without being able to distinguish between doublons or holes (parity projection) \cite{Bakr2009, Sherson2010}, or single spin-component observables alone \cite{Parsons2016, Cheuk2016}, thereby severely restricting the potential of quantum gas microscopy, especially for doped systems. Recent studies at full density- and spin-resolution have shown the capability of simultaneous detection of occupation and spin, including the static \cite{Hilker2017, Salomon2019} or dynamic \cite{Vijayan2019} aspects of spin-charge separation in one dimension or magnetic polarons in two dimensions \cite{Koepsell2019}. However the applied technique \cite{Boll2016} typically strongly constrains available lattice geometries, reduces system size and leads to weaker couplings in the system. 

Alternative methods for spin-resolved readout are based on vector light shifts or electronic dark states and can be partly used to overcome such deficiencies \cite{Wu2019, Kwon2017}, however, they were demonstrated for atomic species not available for Fermi-Hubbard experiments. In addition, it would be desirable to extend quantum gas microscopy techniques beyond two dimensions, thus enabling the control and study of multi-layered systems. This is highly challenging, because the high density of the analyzed systems, on scales of the optical resolution, prevents employment of multi-layer readout schemes that are suitable for systems with large lattice spacing and vanishing tunnel coupling between sites \cite{Nelson2007, Barredo2018, Eliasson2019}. For bosons, a first example of such a scheme has been demonstrated \cite{Preiss2015}. 
 
\begin{figure}[!ht] 
\includegraphics[]{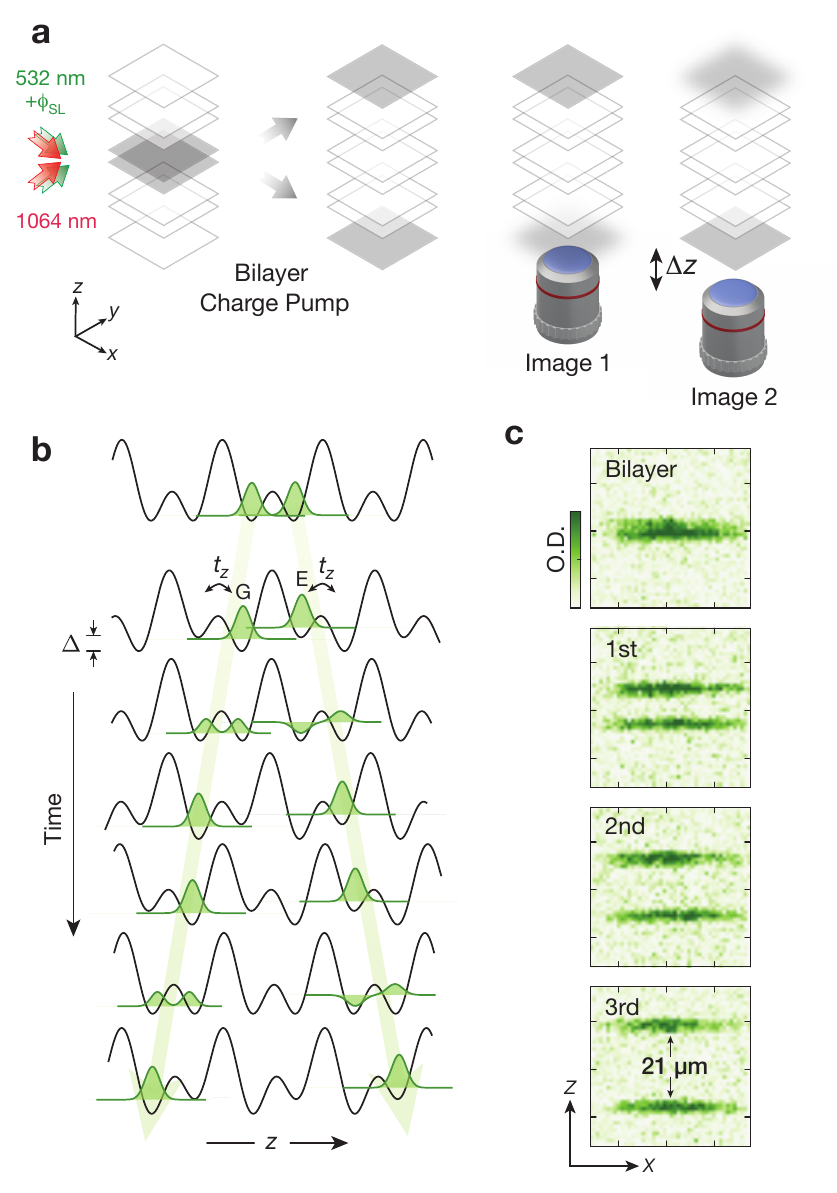}
 \caption{\textbf{Schematics of setup and charge pump.}
 \textbf{a}, We use a bichromatic optical superlattice (red, green) in the vertical direction of our quantum gas microscope to control bilayer systems (gray) and their mutual tunnel couplings $t_{z}$ or energy offsets $\Delta$. The two layers of a bilayer system are pumped to a separation of $21\,\mu$m, which enables subsequent site-resolved microscopy of both layers by shifting the focal plane by $\Delta z$ between the images. \textbf{b}, Charge pumping is achieved by time-dependent modulation of superlattice parameters. Atoms initialized in opposite wells ($G$ and $E$) experience transport in opposite directions. For technical reasons we reset the superlattice phase after each adiabatic passage (see Supplementary Material). \textbf{c}, Absorption images (side view) of an initially coupled bilayer system, whose planes are subsequently pumped in opposite directions for 1,2 and 3 pumping steps, leading to a final separation of 21\,$\mu$m between the two layers.}
 \label{fig:schematic}
 \end{figure}


Here we demonstrate a novel approach that overcomes all these challenges. It allows to realize and image bilayer systems and to obtain full spin- and density-resolved images of two-dimensional quantum gases with arbitrary geometries, including two-dimensional Fermi-Hubbard systems studied here. By using a vertical bichromatic superlattice, we gain full control over coupled layers to implement a  charge pump \cite{Thouless1983,Isart2007,Qian2011, Wang2013,Lohse2016,Nakajima2016}, which makes our scheme especially robust and efficient. We use this quantum pump to separate two initially coupled layers over large distances. The large separation between the layers, far beyond the depth of focus of our imaging system, enables single-site fluorescence imaging of a single layer without significant background contribution from the other layer. By using a vertical magnetic field gradient to split different hyperfine states of atoms into different layers in an initial step, we can use the scheme to implement high-fidelity spin- and density-resolution of two-dimensional Fermi-Hubbard systems. We first benchmark the pumping and bilayer readout technique by imaging the individual site occupations of a bilayer system, consisting of two coupled two-dimensional Mott insulators. Next, we demonstrate full spin- and density-resolution of a single Fermi-Hubbard layer and reveal strong antiferromagnetic spin correlations in the Mott insulating regime.

Our quantum gas microscope realizes spin-$1/2$ Fermi-Hubbard systems using $^{6}$Li in a square optical lattice of spacing $a_{xy}=1.15\, \mu$m and tunnel coupling $t$ in the $xy$-direction \cite{Omran2015}. In the vertical $z$-direction the atomic system is usually confined to a single layer of a highly stable bichromatic optical superlattice (see Fig.\,\ref{fig:schematic}a and Supplementary Material). This vertical lattice exhibits short (long) lattice spacings of $a_{z}^{s}=3\, \mu$m ($a_{z}^{l}=6\, \mu$m) and is created by interfering two laser beams of wavelength $\lambda^{s}=532\,$nm ($\lambda^{l}=1064\,$nm) at an angle of $5.1 ^{\circ}$. The phase difference $\phi_{SL}$ between the two $532\,$nm lattice beams is controlled by shifting the frequency of the $532\,$nm light, thereby enabling full and dynamical control of the resulting superlattice potential (see Supplementary Material). Strongly coupled bilayer systems with tunnel couplings of up to $t_{z}/h=571(1)\,$Hz between two layers
can be realized with typical lattice depths of $V^{s}=11\,E_{R}^{s}$ and $V^{l} = 100\,E_{R}^{l}$, where $E_{R}^{i}$ denotes the respective recoil energy.

An important feature of time-modulated superlattices is the existence of two distinct bands $G$ and $E$ with opposite Chern numbers \cite{Thouless1983, Atala2013, Lohse2016}. These bands cause transport in opposite directions upon the same adiabatic passage of the double-well tilt $\Delta$. This is referred to as geometric pumping and lies at the heart of topological charge pumping in time-modulated superlattices \cite{Qian2011, Wang2013, Isart2007, Lohse2016, Nakajima2016}. As depicted in Fig.\,\ref{fig:schematic}b, an atom initialized in $G$ can be transferred to a neighbouring well by adiabatically changing the energy offset $\Delta$ between the wells at a constant interwell tunnel coupling $t_{z}$. For the same ramp, an atom in $E$ will be transported to a neighbouring well in the opposite direction. Performing $n$ such adiabatic pumping steps can be used to separate two layers, initially displaced by only $a^{s}_{z}$, over macroscopic distances of $a^{s}_{z}+n a^{l}_{z}$. In Fig.\,\ref{fig:schematic}c we show absorption images of atoms initialized in neighbouring layers of the vertical superlattice. Initially, the distance between the two layers is unresolved, however, after three adiabatic pumping cycles we are able to clearly resolve the wide separation of $21\,\mu$m. Optimized pump parameters for high-fidelity transport are described in the Supplementary Material.

\begin{figure}[htb]
\includegraphics[width=0.47\textwidth]{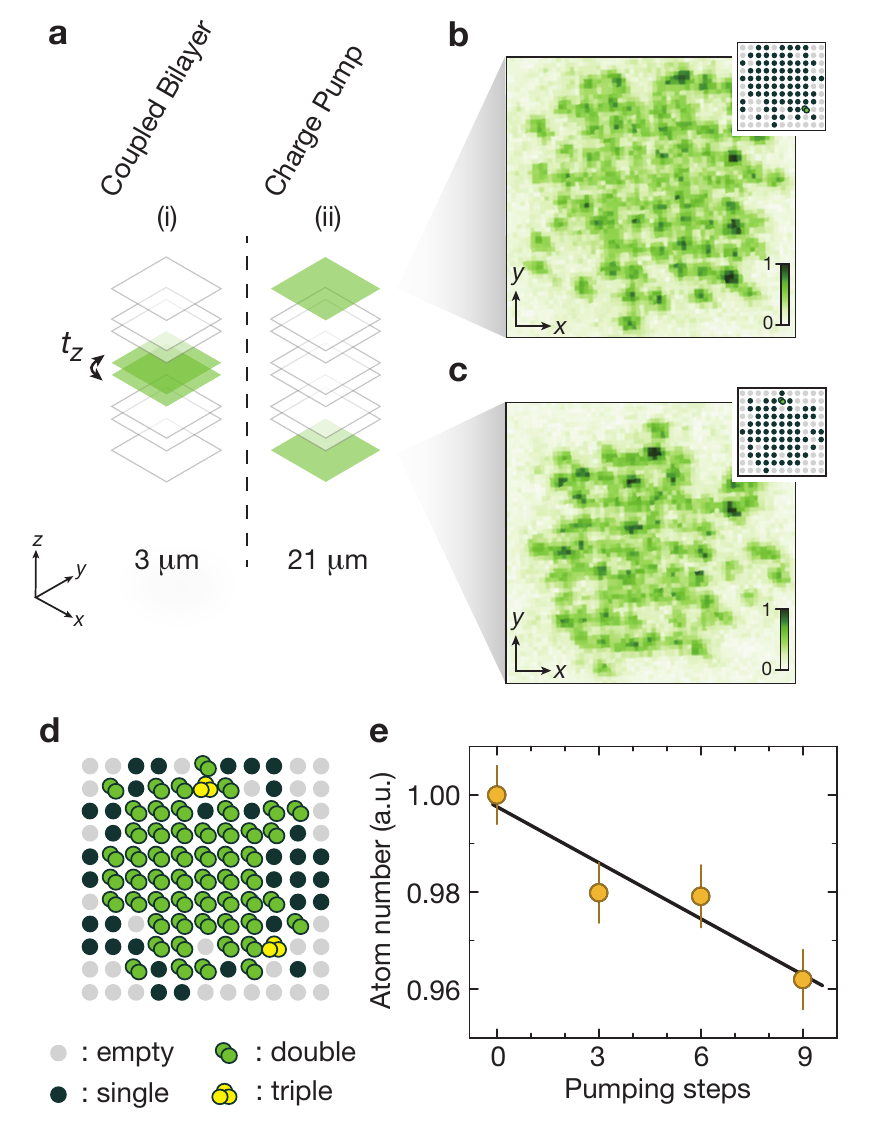}
\caption{\textbf{Bilayer readout and charge pump fidelity.} \textbf{a}, An initially coupled Mott insulating bilayer system is separated over a distance of 21\,$\mu$m using three charge pumping steps. \textbf{b,c}, Single-site resolved fluorescence images and reconstruction (inset) of the respective site occupations of the two separated layers. The images were obtained by shifting the high-resolution objective to sequentially image the two layers. \textbf{d}, Summed occupation of vertically combined  sites (super-sites), relevant for bilayer systems. Here, the two vertically overlapping Mott insulators show a large region of double occupation on a super-site. \textbf{e}, Averaged and normalized number of atoms in a monolayer system as a function of pumping steps. A fit to the data (black line) yields a transfer-fidelity for each pumping step of $\eta =0.996(1)$. Error bars denote one s.e.m..}
\label{fig:bilayerimaging}
\end{figure}

\begin{figure}[htb]
\includegraphics[width=0.47\textwidth]{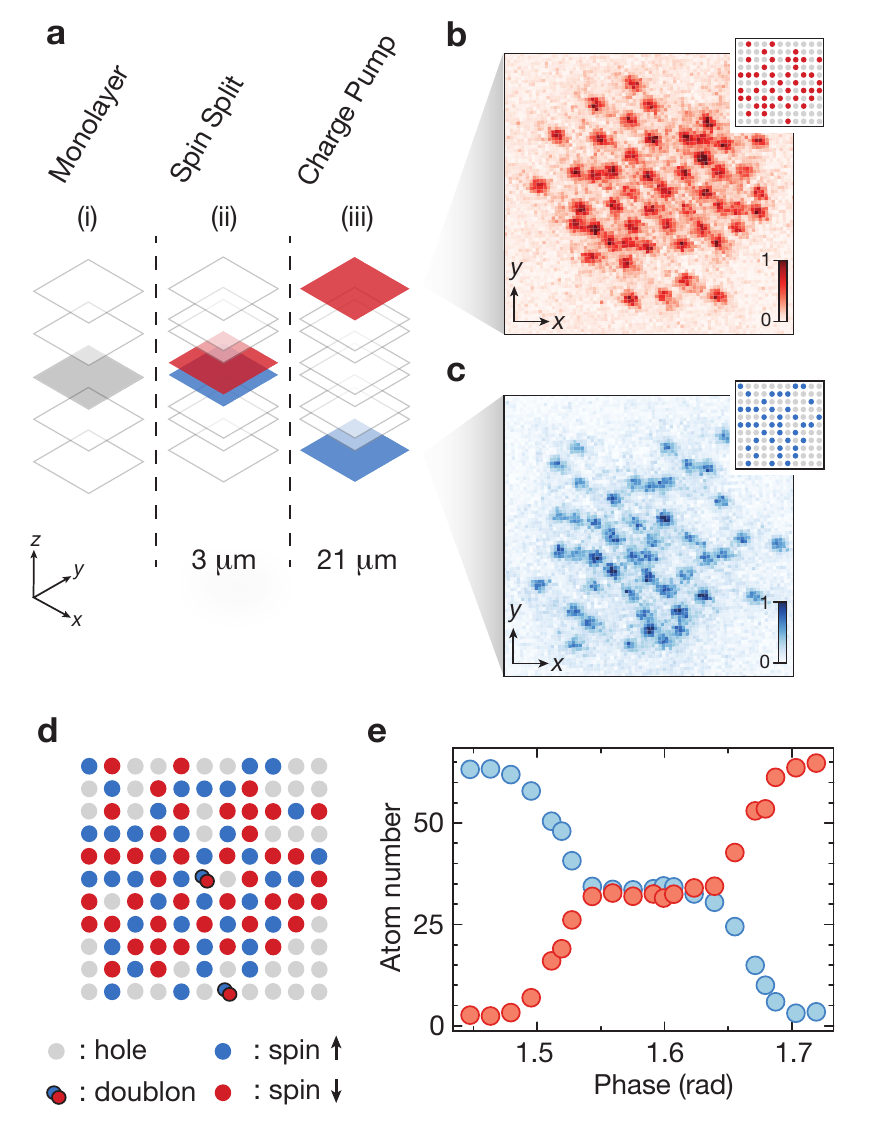}
\caption{\textbf{Spin resolution of two-dimensional systems. a,} The two different spin states of a monolayer spin mixture (here a two-dimensional Fermi-Hubbard system) are split into different layers (red/blue) of a bilayer system with a vertical magnetic gradient. Subsequently, charge pumping and bilayer readout is applied with single-site resolution. \textbf{b,c}, Snapshot of the two spin components of a two-dimensional Fermi-Hubbard system after separation and the full spin and density reconstruction of the original single layer in \textbf{d}. The correct recombination of the spin-layers is ensured by the unique minimization of reconstructed doublons (see Supplementary Material). \textbf{e},  Number of atoms detected in the upper (red) or lower (blue) layer as a function of the superlattice splitting phase. The magnetic field gradient during the initial splitting process creates a broad plateau of superlattice phases $\phi_{SL}$, for which a balanced spin-mixture is successfully split into its constituents. 
}
\label{fig:spinimaging}
\end{figure}


A large separation between planes enables the independent microscopy of each layer. Our imaging system (NA=$0.5$) exhibits a depth of focus below $3\,\mu$m. When taking a fluorescence image with one layer in focus, the other layer will be out-of-focus and only contribute a weak and homogeneous background. After capturing the first image, we shift the focal plane of our imaging system to take an image of the other layer. The parasitic background of the opposite layer is extremely weak, such that no further image processing is necessary and occupations can be reconstructed with high fidelity using our usual deconvolution algorithm \cite{Boll2016}. To demonstrate the independent readout of two layers, we initialize and image a weakly coupled ($t_{z}/t = 1.3$, $\Delta=0$) Mott-insulating bilayer system. After freezing the atomic occupations, three pumping steps are applied to separate the atomic planes. Then we take fluorescence images of both layers in a dedicated pinning lattice \cite{Omran2015} and reconstruct the single-site resolved occupations (see Fig.\,\ref{fig:bilayerimaging}). Furthermore, the combined occupation between vertically neighbouring sites (forming a super-site) is highly relevant for bilayer systems, as a suppression of density fluctuations on such a super-site is expected for band-insulating \cite{Golor2014} or dimer phases \cite{Chen2013}. This observable is now readily accessible (see Fig.\,\ref{fig:bilayerimaging}d) as well as three-dimensional charge correlators to study bilayer physics in the future.

In order to determine the probability of adiabatic transfer per pumping step $\eta$, we track the number of atoms initialized in a single layer for a variable amount of pumping steps $n \in [0,3,6,9]$. We reverse the pump direction after the third pump to avoid leaving the cooling region of our pinning lattice. Fitting an exponential $\eta^ {n}$ to this curve yields a very high pumping efficiency of $\eta=0.996(1)$, underlining the robustness of the separation process. The main limitation of this fidelity is caused by the residual harmonic confinement along the vertical direction of order $\simeq 300\,$Hz, caused by the $xy$-lattices. It detunes distant double-well structures appreciably compared to the superlattice energy scales $t_{z}$ and reduces the fidelity of pumping in both directions.

\begin{figure}[htb]
\includegraphics[]{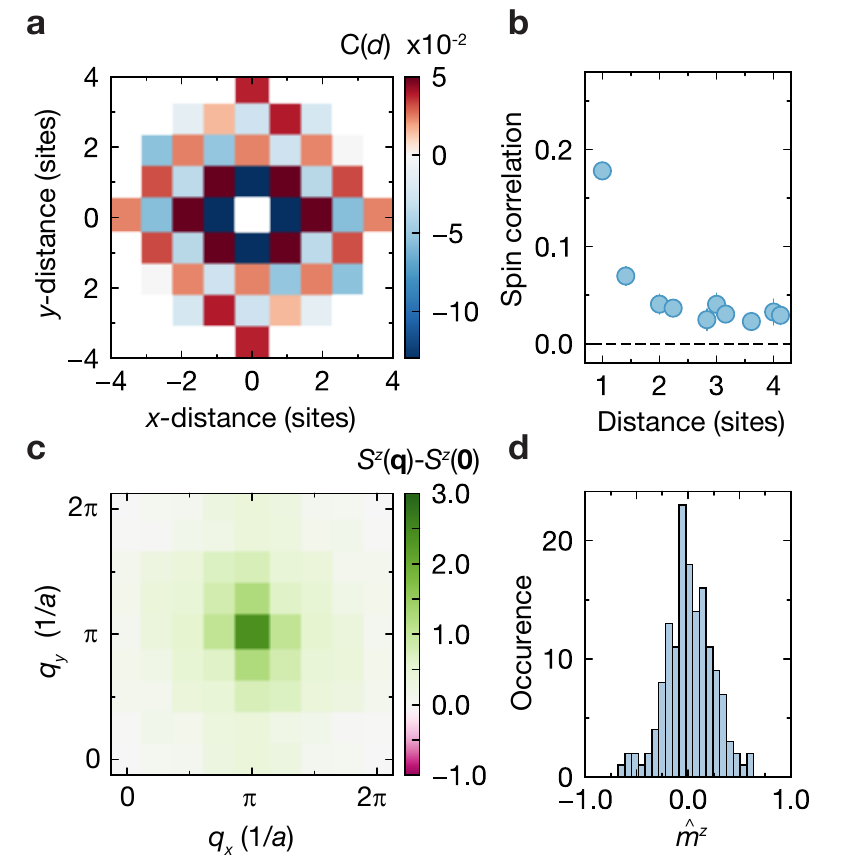}
\caption{\textbf{Two-dimensional spin correlations}, \textbf{a} Averaged spin correlations versus distance of undoped two-dimensional Fermi-Hubbard systems at $U/t=9.3(4)$ obtained by spin- and density-resolved imaging. Sites with density $n < 0.84$ are excluded from the analysis. \textbf{b}, Sign-corrected spin correlations versus radial distance. Error bars denote one s.e.m. \textbf{c}, Spin structure-factor, exhibiting a strong peak at the antiferromagnetic momentum $(\pi,\pi)$. \textbf{d}, Full counting statistics of the staggered magnetization $\hat{m}^{z}$. The figure is based on $150$ experimental realizations.
 }
\label{fig:spincorrelations}
\end{figure}

The bilayer readout scheme presented above can be readily extended to obtain full spin- and density-resolution of a single 2D Fermi-Hubbard plane. After realizing such a single two-dimensional system, we first freeze the motion of the atoms in a deep $xy$-lattice and confine them in a single layer of the large-scale vertical lattice. A strong vertical magnetic field gradient of $45\,$G/cm is then applied to pull the two spins in opposite directions and the short spaced vertical lattice is turned on adiabatically with a phase $\phi_{SL}$ set to obtain symmetric double wells ($\Delta=0$) along the vertical direction. We thereby realize a controlled Stern-Gerlach separation \cite{Boll2016} of the two spin components into two adjacent planes of the bilayer optical lattice (see Fig.\,\ref{fig:spinimaging}). For our experiment, we employ a magnetic offset field below $13\,$G to maximize the differential magnetic moments between the lowest hyperfine states of $^{6}$Li during the splitting process. Once the spin states are split into different layers, we separate the two planes over larger distances using the bilayer readout sequence introduced above. We measure the occupation in each layer as a function of the superlattice phase $\phi_{SL}$ during the splitting to demonstrate the accuracy of the spin splitting (see Fig.\,\ref{fig:spinimaging}e). Working with a spin-balanced system, successful spin-splitting manifests itself as a broad range of superlattice phases $\phi_{SL}$, for which the occupation in both layers is constant. The width of this plateau is determined by the strength of the magnetic field gradient and the lattice parameters along the $z$-direction. Combining the spin-resolved occupations of each layer, we obtain the full density and spin reconstruction of the two-dimensional system (see Fig.\,\ref{fig:spinimaging}d). The excellent agreement between densities and density fluctuations of unpumped and spin-resolved pumped layers excludes the presence of any significant transverse motion caused by the vertical splitting and pumping (see Supplementary Material).

The most stringent benchmark, however, for our spin-resolution technique is the measurement of spin-correlations in a two-dimensional system. For undoped Mott insulators, strong antiferromagnetic correlations are expected to arise in the system in the regime of large on-site interaction versus tunneling $U/t$ and for low enough temperatures, due to the antiferromagnetic Heisenberg spin couplings $J$. We take spin- and density-resolved measurements of an undoped system at an in-plane lattice depth of $6.9(1)\,E_{R}^{xy}$ and $U/t=9.3(4)$ with around $90$ atoms. To ensure perfect recovering of the two-dimensional parent system from the spin-layers, we match both layers based on the unique minimization of reconstructed doublons (see Supplementary Material). At our coldest temperatures, we indeed observe strong antiferromagnetic correlations for our system extending over multiple sites (see Fig.\,\ref{fig:spincorrelations}a).
From the data, we extract spin correlations $C(d)=\langle \hat{S}^{z}_{i}\hat{S}^{z}_{i+d}\rangle$ as well as reveal a strong peak in the corresponding spin structure factor $S(\mathbf{q})$ at the antiferromagnetic quasimomentum $\mathbf{q}=(\pi,\pi)$. We find a mean staggered magnetization of $m^{z}=\sqrt{\langle (\hat{m}^{z})^{2}\rangle }=0.19(1)$ in our system (see Supplementary Material). Comparing with Quantum Monte Carlo calculations of \cite{Mazurenko2017} at $U/t=7.2$, these correlations correspond to a temperature of $k_{B}T\simeq 0.6\,J$, where $k_{B}$ is the Boltzmann constant. Such temperatures are among the lowest that have been reported for ultracold fermionic systems and underline that the detection process employed preserves the intricate correlations in the system.

Our work underlines the unique detection possibilities afforded by versatile and highly controlled superlattice setups for quantum gas microscopy. Next to realizing tunable bilayer systems with independent density readout, we showed how our scheme can be used as a universal technique to obtain full spin- and density-resolution of single 2D Fermi-Hubbard layers for arbitrary 2D lattice geometries. The methods demonstrated here combine robust charge pumping and efficient Stern-Gerlach separation to provide an entirely new degree of control and readout for quantum gas microscopes. Our work extends the capabilities of such systems to coupled layered systems, which are highly relevant in the context of high-$T_{c}$ superconductivity \cite{Maier2011, Liechtenstein1995, Bulut1992, Golor2014, Okamoto2008}. An extension of the scheme to spin detection of a bilayer system requires four images and can be implemented in the future. Furthermore, advanced cooling schemes for two-dimensional systems based on dynamical disentangling of layers are now within reach \cite{Kantian2016} and schemes for non-trivial observables like entanglement entropy \cite{Pichler2013, Islam2015} or angle-resolved-photoemission-spectroscopy \cite{Bohrdt2018, Brown2019} can be realized on the repulsive side ($U>0$) and for higher dimensional systems, based on the technique reported here. 


\bigskip
\begin{acknowledgments}
\textbf{Acknowledgments:} This work was supported by the Max Planck Society (MPG), the European Union (FET-Flag 817482, PASQUANS), the Max Planck Harvard Research Center for Quantum Optics (MPHQ) and under
Germany's Excellence Strategy  -- EXC-2111 -- 39081486. J.K. gratefully acknowledges funding from Hector Fellow Academy. We would like to thank Michael H{\"o}se for early contributions to the superlattice setup.
\end{acknowledgments}
\bigskip


\newpage

\section*{Supplementary Material:}
\setcounter{figure}{0}
\renewcommand\thefigure{S\arabic{figure}}  
\subsection{Superlattice Setup}
As a laser source for the $1064\,$nm light of our superlattice we use around $8\,$W from a Mephisto MOPA. $1\,$W of that light is used as seed light for second-harmonic-generation of the $532\,$nm light. The seed light is sent in a  double-pass configuration through two consecutive AOMs (AODF 4225-2, G$\&$H), which have a tailored extremely large bandwidth of about $130\,$MHz. By controlling the frequency drive of the AOMs, we tune the absolute frequency by $410\,$MHz. The remaining $30\,$mW after AOMs is amplified to $40\,$W with a fiber amplifier (Azurlight Systems) and used to generate $12\,$W of $532\,$nm light in a single-pass second-harmonic generation with a periodically-poled crystal (PPMgSLT, OXIDE). The two beams, which are interfered for the vertical lattice, originate from a single overlapped bichromatic beam. For optimal phase stability, the single overlapped beam is split into two bichromatic beams and aligned onto the atoms, all in a dedicated evacuated and temperature-stabilized chamber. A path length difference of $\Delta L=56\,$cm between the two bichromatic beams in the evacuated chamber allows control of the relative superlattice phase $\phi_{SL}$ without moving physical parts, by shifting the frequency of the $532\,$nm light. Elliptical beam shaping is used to guarantee a low in-plane harmonic confinement at the position of the atoms. 

\subsection{Superlattice Control}
\begin{figure}[ht]
\includegraphics[width=0.37\textwidth]{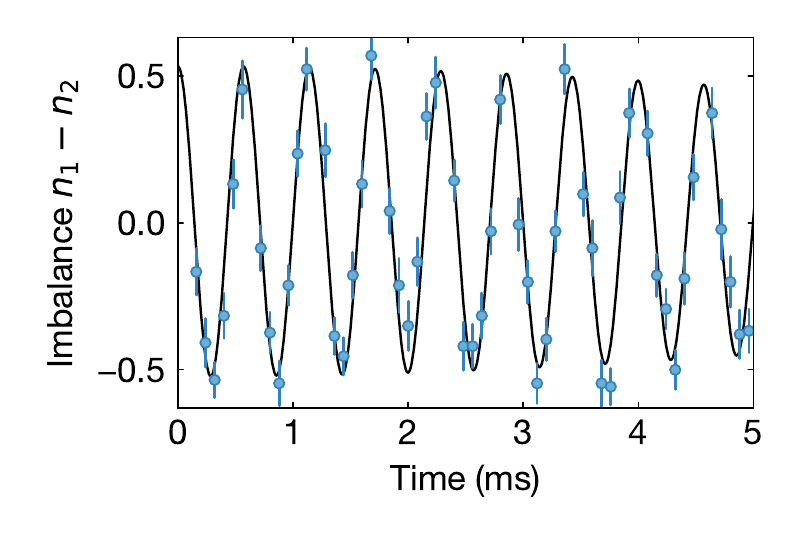}
\caption{\textbf{Rabi oscillation between two layers.} Imbalance of the local densities $n_{1}$ and $n_{2}$ of two layers after quenching on inter-layer tunneling, averaged over a region of three sites. By fitting the observed Rabi frequency (solid black), we extract the interlayer tunneling strength of $t_{z}=h\times 571(1)\,$Hz for vertical lattice depths of $11\,E_{R}^{s}$ and $100\,E_{R}^{l}$. Error bars denote one s.e.m.}
\label{fig:rabi}
\end{figure}

The optical superlattice potential is given by $V=V_{s}\cos^2 (k_{s}z+\phi_{SL})+V_{l}\cos^2 (k_{l}z+\pi/4)$, with lattice depths $V_{s,l}$ , wave vectors $k_{s,l}=\pi/a_{s,l}$ and superlattice phase $\phi_{SL}$. Due to the arm length difference of the two interfering beams of the short lattice, a change in frequency changes not only $k_{s}$, but also $\phi_{SL}$. As the frequency change is small compared to the absolute frequency of the light, the wavevector change is irrelevant. The superlattice phase, on the other hand, is tunable by an amplitude of $1.6\,\pi$ with $\Delta \nu=820\,$MHz.
\\
We verify the tunnel coupling and coherence between two layers by measuring Rabi oscillations. A Mott insulating state was prepared at $26\,E_{R}^{xy}$ in an isolated single layer and quenched to a strong coupling with its neighbouring layer. After quenching off the coupling, we apply bilayer readout and measure the single-site resolved occupation per layer as a function of the coupling duration. In Fig. \ref{fig:rabi}, the oscillation of the occupation is shown for a quench to $V_{s}=11\,E_{R}^{s}$ and $V_{l}=100\,E_{R}^{l}$. We extract a tunnel coupling of $571(1)\,$Hz for those parameters by an exponentially decaying sinusoidal fit, where the error denotes the uncertainty of the fit.

\subsection{Pump Sequence}
Our range of available superlattice phases contains two distinguishable symmetric double-well configurations at $-\pi/2$ and $\pi/2$. We realize a coupled bilayer system or the splitting of a monolayer into its spin components at a symmetric superlattice phase of $\pi/2$. During the application of splitting and threefold pumping, the atoms are frozen in the $xy$-plane with $43\,E_{R}^{xy}$ and vertically with $49\,E_{R}^{s}$ and $100\,E_{R}^{l}$ for a duration of $400\,$ms. To prepare the first pump, $\phi_{SL}$ is ramped within $5\,$ms from $\pi/2$ to $-\pi/2+\delta\phi/2$, where $\delta \phi=0.15\,\pi$. Then the double-well tunnel coupling $t_{z}$ is turned on by ramping the short lattice down to $11\,E_{R}^{s}$ in $20\,$ms. The first pump is performed by ramping the superlattice phase within $3\,$ms by $\delta \phi$ across the symmetric configuration to a final value of $\phi_{SL}=-\pi/2-\delta \phi/2$. During this phase sweep the transport of atoms to the neighbouring layer happens. Eventually, the short lattice is ramped back to $49\,E_{R}^{s}$ and the next pump is performed at $\phi_{SL}=(-1)^{\chi+1}\,\pi/2$, where $\chi$ is the number of already applied pumps. These parameters were optimized for highest overall pump fidelity.

\subsection{Bilayer Readout} 
\begin{figure}[ht]
\includegraphics[width=0.4\textwidth]{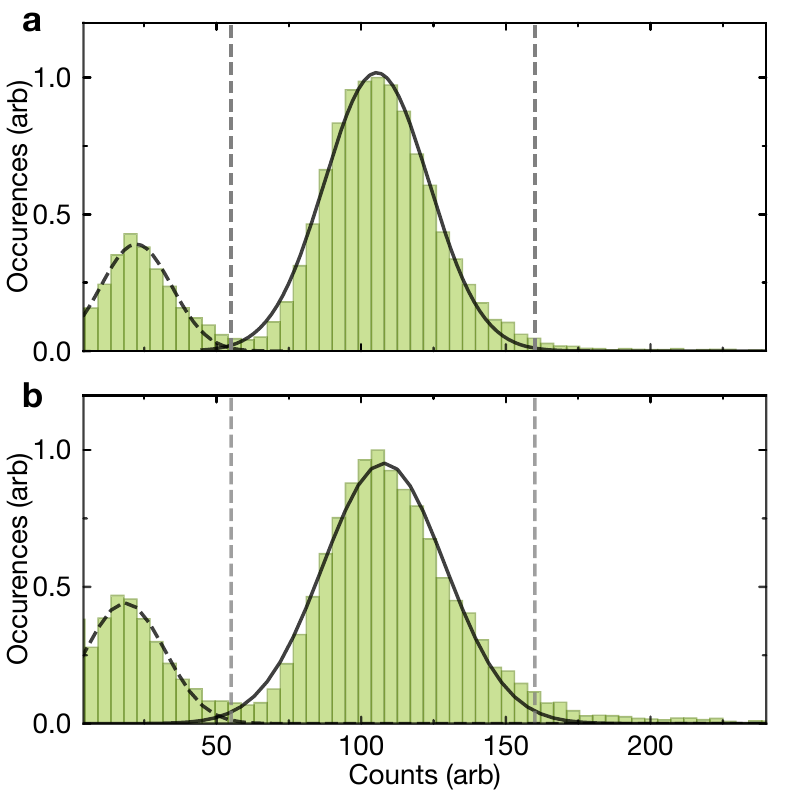}
\caption{\textbf{Bilayer reconstruction.} Histogram of counts per lattice site in layer one (\textbf{a}) and two (\textbf{b}). Solid and dashed black lines are gaussian fits to the peaks of single and zero occupation. Vertical dashed grey lines denote the threshold for single and double occupation.}
\label{fig:hist}
\end{figure}

The bilayer system was prepared from a harmonically confined atom cloud, held in a single layer of the large-spaced vertical lattice. At a fixed balanced superlattice phase ($\Delta=0$), the in-plane lattice potential and short-spaced vertical lattice were ramped simultaneously to a depth of $V_{xy}=11\,E_{R}^{xy}$, $V_{s}=19\,E_{R}^{s}$ with a final on-site interaction $U$ versus in-plane tunneling $t$ of $U/t \simeq 50$. The motion of atoms was then quenched rapidly, by ramping to a lattice depth of $V_{xy} \simeq 40\,E_{R}^{xy}, V_{s} \simeq 50\,E_{R}^{s}$.
The deconvolution of the fluorescence images obtained from each layer was done with a Lucy-Richardson algorithm implemented in our previous work \cite{Boll2016}. The resulting histograms of counts per lattice site are shown in Fig. \ref{fig:hist}. By fitting a gaussian to the zero- and single-occupation peaks, we determine a fidelity to distinguish holes from singlons by summing the probability for false negatives and false positives. For the reconstruction of a bilayer system of two Mott-insulator, this fidelity is $99.4\,\%$ for layer one and $99.0\,\%$ for layer two. For presentation purposes of individual snapshots, we subtracted a weak gaussian background from the pumped images presented in the main text. This is not used nor required for any reconstruction algorithm and is merely used to obtain a more homogeneous image for the eye.

\subsection{Spin Reconstruction}
\begin{figure}[ht]
\includegraphics{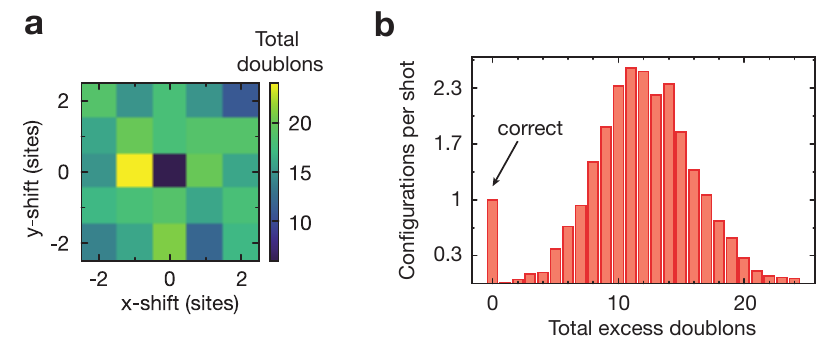}
\caption{\textbf{Centering of two layers for spin reconstruction.} To accurately center the two spin layers, one layer is shifted versus the other by up to $\pm2$ lattice sites in $x$- and $y$- direction and the reconstructed total doublon number is compared. \textbf{a}, Total number of reconstructed doublons for the $5\times5=25$ different shift configurations of a single experimental shot. Here, the configuration with zero shifts applied uniquely minimizes the number of doublons. \textbf{b}, Average number of configurations grouped by the total number of doublons in excess of the minimum number found. There always exists exactly one correct shift configuration, which contains a minimum number of doublons.}
\label{fig:centering}
\end{figure}

To combine the reconstruction of the two spin-layers to obtain the reconstructed spin and density information of the two-dimensional parent system, the right lattice sites along the vertical direction had to be paired. The vertical movement of our objective by $21\,\mu$m between two images sometimes caused small movements of the objective in the transverse direction. This resulted in shot to shot fluctuation of the displacement of the two images on the order of up to one lattice site. To pair the sites correctly, we compared $25$ pairing possibilities, given by shifting one layer in $x,y$ direction by $\pm2$ lattice sites. Since our two-dimensional systems have large Mott-insulating regions with unit occupation, we found the right configuration was uniquely given by the lowest amount of reconstructed double occupations. This can be seen in Fig.\,\ref{fig:centering}a, which shows the number of doublons of all $25$ shift configurations for a single experimental shot. As can be seen from the average distribution of doublons found in the shift configurations (see Fig.\,\ref{fig:centering}b), there exists exactly one unique shift configuration for all shots.  In $12.7\, \%$ of the data the layers were centered with non-zero shifts. This centering of  layers can be mitigated in future experiments by a second imaging path with different focus position, which does not require a physical displacement of the objective.

\subsection{Density Fidelity of Spin Readout}
\begin{figure}
\includegraphics{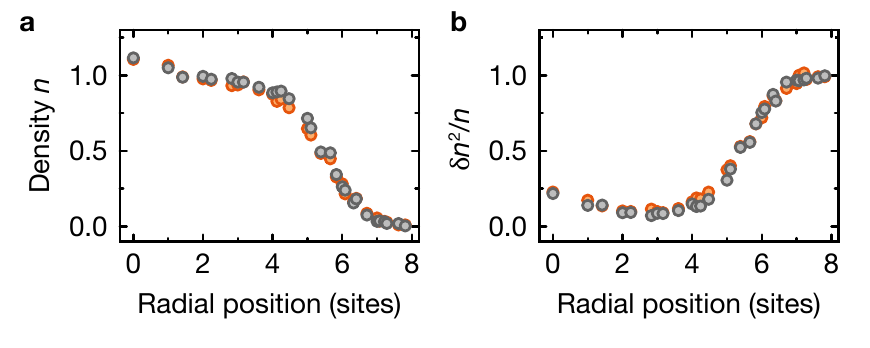}
\caption{\textbf{Comparison of methods.} Averaged radial density (\textbf{a}) and fluctuation profiles (\textbf{b}) of an identical Mott-insulating system measured without (black) and with (red) our spin-resolved technique. Error bars denote one s.e.m. and are smaller than the point size.}
\label{fig:density}
\end{figure}

In order to show no particle motion occurs during spin-splitting and pumping, we compared the two-dimensional density and its fluctuations measured with our spin-readout to an identical un-splitted and un-pumped system. As shown in Fig. \ref{fig:density}, the density and its normalized fluctuations are in good agreement for the two methods.

\subsection{Spin Correlations}
Undoped two-dimensional Fermi-Hubbard systems were prepared from a harmonically confined cloud by ramping the in-plane lattice depth to a final value of $6.9(1)\,E_{R}^{xy}$ within $100\,$ms. The final scattering length was set to $810\,a_{B}$, where $a_{B}$ denotes the Bohr radius and the system was vertically confined in a strongly tilted superlattice ($\phi_{SL}=0$) of depths $V_s = 50\,E_{R}^{s}$ and $V_{l}=100\,E_{R}^{l}$. Spin correlations $\langle \hat{S}^{z}_{i}\hat{S}^{z}_{j} \rangle$ are evaluated on sites with mean density above  $0.84$, which comprises $N=47$ lattice sites. For the spin structure factor, we pad the $N$ sites of each snapshot with zeros to a $9\times9$ system. Then we compute $S^{z}(\textbf{q})=  \langle \sum_{i}\hat{S}^{z}_{i}e^{-i\textbf{q}\textbf{R}_{i}} \times \sum_{j}\hat{S}^{z}_{j}e^{i\textbf{q}\textbf{R}_{j}}\rangle/(\sum_{k}\langle \hat{S}^{z}_{k}\hat{S}^{z}_{k}\rangle)$. The staggered magnetization is computed according to $\hat{m}^{z}= \sum_{i}(-1)^{i}\hat{S}^{z}_{i}/(SN)$ with $S=1/2$. To evaluate the mean and its standard error of $m^{z}$, we performed a bootstrap by resampling our data $200$ times with replacement.

\bibliography{bibliography}
%
%
%

\end{document}